\documentclass{article}

\usepackage{arxiv}

\usepackage[utf8]{inputenc} 
\usepackage[T1]{fontenc}    
\usepackage{hyperref}       
\usepackage{url}            
\usepackage{booktabs}       
\usepackage{amsfonts}       
\usepackage{nicefrac}       
\usepackage{microtype}      
\usepackage{graphicx}
\usepackage[square,sort,comma,numbers]{natbib}
\usepackage{doi}

\usepackage{mathpartir}
\usepackage{listings}
\usepackage{array,multirow,amsmath,amssymb,graphicx,array,url} 
\usepackage{xcolor}
\usepackage{float}
\usepackage{multirow}
\usepackage{subcaption}
\DeclareMathOperator*{\argmax}{arg\,max}

\newcommand{\matr}[1]{\mathbf{#1}}

\title{Predicting affinity ties in a surname network}

\author{Marcelo Mendoza \\
	Department of Computer Science \\
	Pontifical Catholic University of Chile \\
	Santiago, Chile \\
	\texttt{marcelo.mendoza@uc.cl} \\
\And
    Naim Bro \\
	School of Government \\
	Universidad Adolfo Ib\'a\~nez \\
	Santiago, Chile \\
	\texttt{naim.bro@uai.cl} \\
}

\renewcommand{\headeright}{Preprint version}
\renewcommand{\undertitle}{Preprint version}
\renewcommand{\shorttitle}{Predicting affinity ties in a surname network}

\hypersetup{
pdftitle={Predicting affinity ties in a surname network},
pdfauthor={Marcelo Mendoza, Naim Bro},
pdfkeywords={Affinity networks},
}

\begin{document}
\maketitle

\begin{abstract}
	From administrative registers of last names in Santiago, Chile, we create a surname affinity network that encodes socioeconomic data. This network is a multi-relational graph with nodes representing surnames and edges representing the prevalence of interactions between surnames by socioeconomic decile. We model the prediction of links as a knowledge base completion problem, and find that sharing neighbors is highly predictive of the formation of new links. Importantly, We distinguish between grounded neighbors and neighbors in the embedding space, and find that the latter is more predictive of tie formation. The paper discusses the implications of this finding in explaining the high levels of elite endogamy in Santiago.
\end{abstract}


\section{Introduction}

In countries where citizens inherit both parents' last names, surnames are a source to uncover the population structure of society. In previous research \cite{Bro:21}, we built a network from surname pairs that revealed strong patterns of socioeconomic segregation in Santiago, Chile. In this study, we model the formation of affinity ties between surnames, and find that the formation of new links is highly predictable.

The method predicts links as a knowledge base completion problem. We evaluate several state-of-the-art knowledge base completion models and find that TuckER, which is based on a three-way tensor decomposition, obtains the best results.

The findings bear implications for the literature that examines the mechanisms behind group segregation. Traditional explanations focus on homophily, the propensity of people to connect to others of their own kind -- the "birds of a feather" mechanism~\cite{goodreauBirdsFeatherFriend2009}. Advances in network statistical modelling in the last decade have disentangled homophily from other, network-specific mechanisms that help segregate groups~\cite{wimmerRacialHomophilyERG2010}. People not only tend to prefer connecting with their in-groups, they are also prone to close triangles. This is the notion of triadic closure or "friend-of-a-friend" (FOAF) mechanism: A and B are likelier to become friends if they have C as a common friend ~\cite{granovetter73,burtStructuralHolesSocial1995,snijdersIntroductionStochasticActorbased2010,deutschlandSociologyInquiriesConstruction2009}. FOAF independently exacerbates the separation of groups, because the friend-of-a-friend is often an in-group.

Our experiments show that triadic closure operates not only on the empirical network. We compute the embeddings of all nodes, and find that those that share a common neighbor in the \emph{embedding space} are very likely to connect. Moreover, proximity in the embedding space accounts for more than half of the triadic closures, and it is a stronger predictor of link formation than proximity in the empirical network. Sharing a similar location in the embedding space can be thought of as being structurally similar.

The substantive implication is that class divisions in Chile do not occur only because people prefer connecting and marrying others of their own socioeconomic status, but because they connect to others occupying similar locations in the network of interactions.

Along with the publication of this study, we release the surname affinity network SA19k (\textbf{S}urname \textbf{A}ffinity \textbf{19k} entities), a knowledge base for the prediction of surname affinity ties in Santiago, Chile. This is a multi-relational graph, where nodes are surnames and links are affinity relations divided by decile in the socioeconomic distribution.

\section{Materials and methods}

\subsection{Data}

We use the data introduced in Bro \& Mendoza \cite{Bro:21}. This dataset is based on the Chilean electoral registry of 2012, which contains the full name and the address of all individuals eligible to vote for political authorities in Chile. We only use the data of residents in Santiago, Chile, totaling 4,652,933 individuals. We also use the Territorial Well-being Index of 2012 \cite {citcentrodeinteligenciaterritorialIndiceBienestarTerritorial2012}, which indexes the mean socioeconomic status of every census administrative unit down to the block level, totaling 39901 blocks.

Individuals' socioeconomic status was assigned based on the mean socioeconomic level of the blocks where they live. Socioeconomic status was transformed into a 0-100 range using the formula $ z {i} = \dfrac{x {i} -min (x)} {max (x) -min (x)} $, where $ x = ( x {1}, ..., x {n}) $ and $ z {i} $ is the $ i $-th element of the normalized vector. For every person in the list, the final dataset contains the paternal and maternal surnames, an assigned socioeconomic status, and the block identification.

\subsection{SA19k: A paternal-maternal surname knowledge base}

We build a knowledge base that comprises paternal-maternal surname ties and income distributions based on the data described above. To create the knowledge base, we build an undirected network based on paternal-maternal surname affinities, where each node represents a surname. To create the edges, we count the number of individuals who share a pair of surnames, irrespective of the paternal-maternal or maternal-paternal order. Each pair of nodes was connected with ten relations, representing the deciles of the income distribution. The edge's weight represents the number of individuals in each decile of the income distribution. This network has 76649 surnames and 1750802 edges.

Following Mateos \textit{et al.} \cite{mateosEthnicityPopulationStructure2011}, we remove surname pairs if their weight denoted by $n_ {ss}$ is less than a threshold $ \dfrac{k \cdot n_ {s1} \cdot n_ {s2}} {N} $, where $k > 1$ is the security parameter of the method, $n_ {s1}$ is the number of occurrences of the surname $s_1$, $n_ {s2}$ is the number of occurrences of the surname $s_2$, and $N$ is the number of individuals in the sample. In a uni-relational graph, $\dfrac{k \cdot n_ {s1} \cdot n_ {s2}} {N}$ is $k$ times the expected number of co-occurrences of both surnames if the surnames are linked at random. A high value of $k$ generates a sparse network where ties are indicative of affinity. 

Mateos \textit{et al.} \cite{mateosEthnicityPopulationStructure2011} report that $k$ must be tuned to balance reliability and sparseness. We test with $k = 10$, $20$, and $30$, finding that $k = 20$ allows us to obtain a network that does not underrepresent any of the income distribution deciles. In addition, we removed rare surnames with less than twenty occurrences. Next, we removed surnames that fall in the periphery of the network using a $k$-core analysis of the network. The nodes that fall below a threshold $k=2$ were removed, preserving the triangles that form the atomic units of relationships that characterize the network. The resulting network contains 19041 nodes and 187563 edges, with an average degree of 19,7. We named this network \textbf{SA19k} (\textbf{S}urname \textbf{A}ffinity \textbf{19k} entities).

\begin{figure}[t!]
    \centering
    \begin{subcaptionblock}{.49\columnwidth}
        \includegraphics[width=\columnwidth]{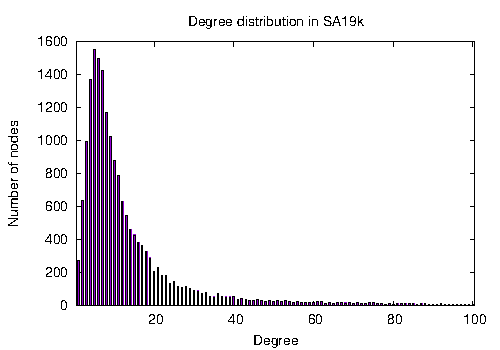}
        \caption{Node degree distribution in SA19k.}
    \end{subcaptionblock}
    \begin{subcaptionblock}{.49\columnwidth}
        \includegraphics[width=\columnwidth]{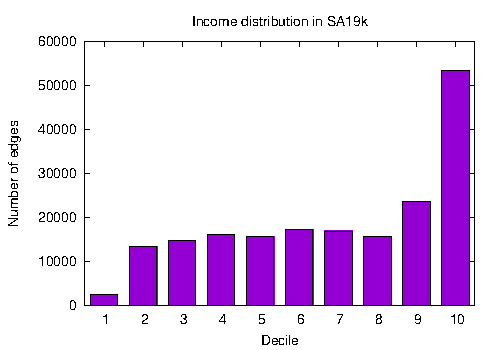}
        \caption{Income decile distribution in SA19k.}
    \end{subcaptionblock}
    \caption{\textbf{Node degree and income decile distribution in SA19k. } a) Even though many nodes have around 5-7 neighbors, the degree distribution shows the presence of hubs; b) Most of the affinity relationships occur in the highest income decile, totaling 29\% of the edges of SA19k. On the other hand, affinities in the lowest decile are almost unexistent. The other income deciles are quite similar, with around 8\% of the edges per decile.}
    \label{fig:1}
\end{figure}

Fig \ref{fig:1} shows the distribution of node degrees and socioeconomic distributions by decile. SA19k's degree distribution exhibits that a few nodes have many connections while most nodes have low connectivity. In specific, most nodes have around 5-7 neighbors (Fig \ref{fig:1}a) and a small number of nodes have more. The income decile with the highest presence in SA19k is the one with the highest income (d10), with almost 29\% of the network nodes (Fig \ref{fig:1}b). The lowest decile has a minor presence, and each one of the deciles 2 to 9 have around 8\% of the network nodes. This shows that surnames at the higher end of the socioeconomic distribution (d9-d10) are more prone to clustering. 

\begin{figure}
    \centering
    \includegraphics[width=\columnwidth]{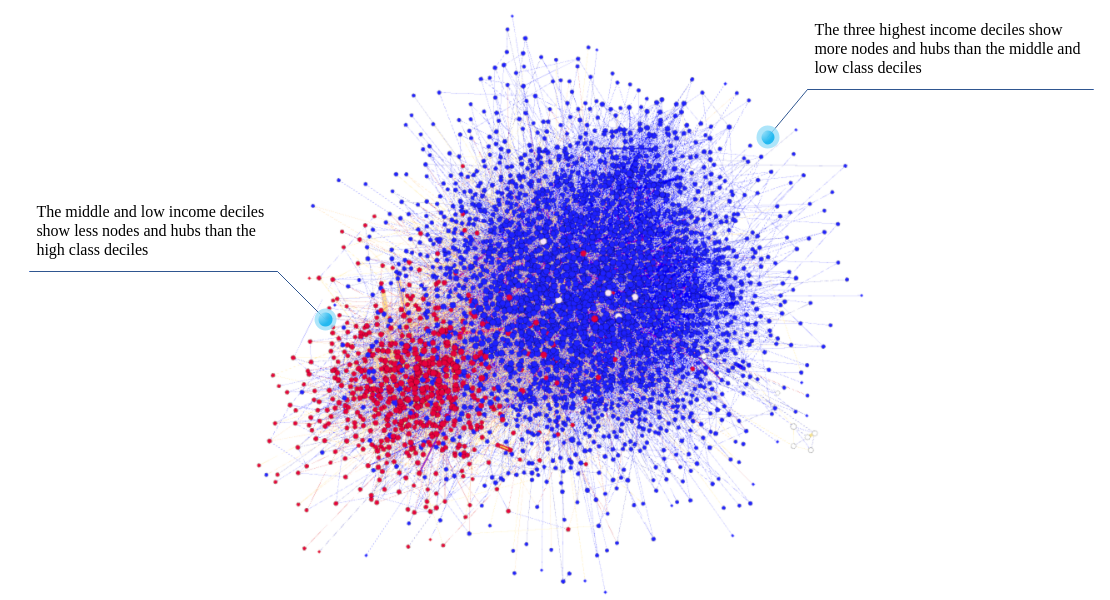}
    \caption{\textbf{A visualization of SA19k.} SA19k is highly modular, with modularity score = 0.802. Low-middle, and high-income deciles are grouped into two modularity-based partitions. The blank nodes do not belong to any of them. Low-middle income distribution deciles (in red color) show less nodes and hubs than the rest of the network. High income deciles (in blue color) show more nodes and hubs than the rest of the network. The size of each node represents its degree.}
    \label{fig:2}
\end{figure}

Fig \ref{fig:2} visualizes SA19k, with colors red and blue corresponding to the main communities identified using Louvain's modularity maximization based community detection algorithm \cite{Louvain:08}. Nodes in the red community are more present in deciles 1-7, and nodes in the blue community belong to deciles 8-10. The network has a high modularity score of 0.802, which increases along with socioeconomic status.

\subsection{Surname affinity prediction model}

We study SA19k using predictive modeling. SA19k can be seen as a knowledge graph for surname affinity conditioned on income distribution. A predictive model based on SA19k can reveal unexpected connections between surnames. Our approach to predictive modeling is based on knowledge graph completion, where a model learns representations of nodes and edges to predict missing links. Link prediction reveals triplet-based facts, where each triplet represents a source and target surname, and each edge corresponds to a decile in the income distribution.

We model SA19k as a knowledge graph as follows. Let $\mathcal{V}$ be a set of nodes representing entities and let $\mathcal{R}$ be a set of relations. We define $\mathcal{E} \subseteq \mathcal{V} \times \mathcal{R} \times \mathcal{V}$ as the set of edges connecting pairs of entities in $\mathcal{V} \times \mathcal{V}$ through a specific relation $r \in \mathcal{R}$. A knowledge graph $\mathcal{G} (\mathcal{V}, \mathcal{E}, \mathcal{R})$ is a set of observed triplets $\langle h, r, t \rangle$, such that $h, t \in \mathcal{\mathcal{V}} \times \mathcal{V}$ and $r \in \mathcal{R}$. The nodes $h, t$ represent head and tail entities connected through $r$. We distinguish between head and tail entities when $r$ is a directed relationship that starts from $h$ and goes to $t$. $\mathcal{G} (\mathcal{V}, \mathcal{E}, \mathcal{R})$ is an undirected knowledge graph if every $r \in \mathcal{R}$ has the same meaning in both directions. In the case of SA19k, we model relationships in a non-directional way, without distinguishing between heads and tails and therefore, using maternal and paternal surnames interchangeably.

Knowledge graph completion is the task of using $\mathcal{G}$ to predict missing facts. Given $r \in \mathcal{R}$, tail prediction addressed the $\langle h, r, ? \rangle$ task, i.e., for a known head entity $h$ and a given relation $r$, we infer from $\mathcal{G}$ the most likely tail node to which $h$ will connect through $r$. Analogously, head prediction is a $\langle?, r, t \rangle$, i.e., predicts the most likely head entity $h$ to connect $t$ through $r$.

Machine learning-based approaches for knowledge graph completion use $\mathcal{G}$ to capture latent features of entities and relations. These features embody the semantic of the original knowledge graph, vectorizing nodes and relationships. For this purpose, the method learns latent representations, known as embeddings, defining a scoring function $\phi$ to estimate the probability of $\langle h, r, t \rangle$ of being observed in the empirical network. The scoring function operates on the embeddings of $h$, $r$, and $t$, represented in vector notation by $\mathbf{h}$, $\mathbf{t}$, $\mathbf{r}$. Then, the method calculates $\phi (\mathbf{h}, \mathbf{r}, \mathbf{t})$, fitting the embeddings to maximize the score of the observed facts, and also minimizing the score of unobserved facts in $\mathcal{G}$.

During the prediction phase, for a given head entity $h$ and a given relation $r$, missing links are inferred determining the entity that gets the highest score according to the scoring function:

$$ \argmax_{e \in \mathcal{V}} \phi(\mathbf{h}, \mathbf{r}, \mathbf{e}).  $$

Similarly, the link prediction task can be defined for head entities, looking for the entity $e \in \mathcal{V}$ that maximizes the scoring function $\phi(\mathbf{e}, \mathbf{r}, \mathbf{t})$ for a fixed $r$ and $t$.

\subsection{Machine learning-based approaches for knowledge graph completion}

To cover the broadest possible range of methods and architectures in the evaluation, we identified representative methods of different model families, taking care that these methods achieve state-of-the-art performances in knowledge graph completion and have open-source implementations that favor the reproducibility of the reported results. We identified three families of models:

\begin{itemize}
\item[--] Tensor decomposition models: These models are based on tensor decomposition techniques of the KG adjacency matrix, which is modeled as a three-dimensional tensor. The tensor representation consists of a set of stacked relationship adjacency matrices. These tensors are decomposed into a combination of low-dimensional vectors used as embeddings for the entities and relations of the KGs. Since tensor decompositions, in general, do not overfit the empirical KG, they are expected to have better generalization capabilities than other methods. We identified three methods of this family that show competitive results in benchmark data \cite{Rossi:21} and that have implementations that allow reproducing their results. These methods are DistMult~\cite{Yang:15} and ComplEx~\cite{Trouillon:16}, both belonging to bilinear type models since they represent the relationship matrix as a bi-dimensional tensor. Another prominent method of this family is the TuckER~\cite{Balazevic:19} method, a non-bilinear method based on Tucker's tensor decomposition~\cite{Tucker:66}.

\item[--] Geometric models: Geometric models interpret relationships as geometric transformations in the latent space of entity embeddings. These methods adjust the parameters of the node embeddings and the geometric transformation function that represents the relations. Given a head embedding, the transformation in the latent space maps it to the tail embedding of the observed fact. The parameters of the embeddings and relationships are adjusted to minimize the distance in the latent space between the head and tail embeddings connected by empirical facts. Geometric models need to verify constraints to ensure that the transformations in latent space are affine. We identified four geometric models representing this family: TransE~\cite{Bordes:13} and TransH~\cite{Wang:14}, based on translational type transformations; CrossE~\cite{Zhang:19}, based on translations with additional embeddings for KG relationships; and RotatE~\cite{Sun:19}, which uses rotations (translations in polar coordinates) to represent relationships between entities in latent space.

\item[--] Deep learning models: These models use layered neural network architectures to learn the embeddings of the KG entities and relationships. The embeddings are learned in conjunction with the weights and biases of each layer of the network, which makes these models gain in expressiveness but assume risks of overfitting. We identified three methods from this family of models that show competitive results in benchmark data \cite{Rossi:21}. ConvE~\cite{Dettmers:18} uses a convolutional network architecture to encode the head entity and relationship embeddings, which produce an outcome in the network output layer that encompasses the KG's set of tail embeddings. The network is trained to maximize the scores of the observed facts, minimizing a categorical cross-entropy loss function. ConvTransE~\cite{Shang:19} reuses this architecture so that the transformation in the latent space between head and tail embeddings corresponds to translations. Finally, SACN~\cite{Shang:19} extends ConvTransE by adding a decoder that reconstructs the observed facts, replacing the loss function with a logit softmax.
\end{itemize}

\subsection{Model selection}

For model selection, we partitioned SA19k into training, validation, and testing folds. We did this by sampling triplets at random. Both the validation and testing partitions have 5000 triplets each, leaving the training partition with 177563 triplets. The proportions between training/validation/testing triplets are typical in knowledge base completion, and very similar to those used to study predictability in the Wordnet database \cite{Dettmers:18}. In S1 Table, we present the results of the validation; with TuckER producing the best statistics. The results in S1 Table show that TuckER obtains the highest precision in the top positions of each list of ranked entities, both in top-1 and top-3 lists. While its top-10 results are competitive but outperformed by RotatE, at MRR, TuckER also outperforms its competitors.

\subsection{Tucker decomposition for link prediction}

The best performing method in SA19K, TuckER~\cite{Balazevic:19}, is a link prediction method based on Tucker's tensor decomposition~\cite{Tucker:66}. The representation of the knowledge graph used by this method corresponds to a tensor (a multidimensional array) produced by the tensor products of $N$ vector spaces. The order of a tensor corresponds to its number of dimensions, 3 in the case of the usual Tucker decomposition. We will denote the input $\langle i, j, k \rangle$ of a three-way tensor $\mathcal{X}$ by $x_{ijk}$.

\begin{figure}
    \centering
    \begin{subcaptionblock}{.49\columnwidth}
        \includegraphics[width=\columnwidth]{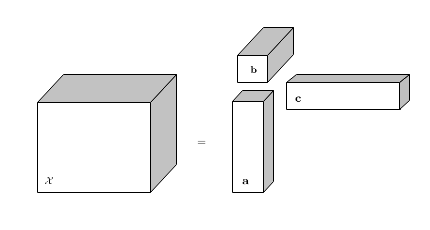}
        \caption{Three-way tensor.}
    \end{subcaptionblock}
    \begin{subcaptionblock}{.49\columnwidth}
        \includegraphics[width=\columnwidth]{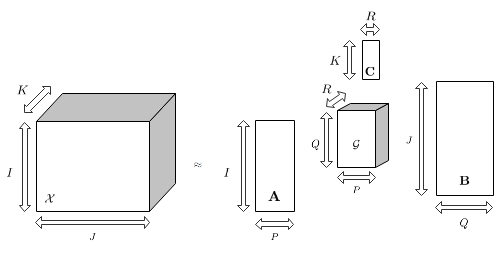}
        \caption{Three-way Tucker decomposition.}
    \end{subcaptionblock}
    \caption{\textbf{Three-way tensors.} a) Three-way tensors are produced by the tensor products of 3 vector spaces; b) The Tucker decomposition factorizes a three-way tensor in a core tensor and three matrices.}
    \label{fig:3}
\end{figure}

We introduce some definitions needed to define the Tucker decomposition properly. An $n$-way tensor $\mathcal{X}^{I_1 \times \ldots \times I_n}$ is rank-one if it can be expressed by the outer product of $N$ vectors: $\mathcal{X} = a^{(1)} \circ a^{(2)} \circ \ldots \circ a^{(n)}$, where $a^{(i)}$ represents the $i$-th vector of a sequence. Note that each entry of $\mathcal{X}$ can be written as $x_{i_1 i_2 \ldots i_n} = a_{i_1}^{(1)} \cdot a_{i_2}^{(2)} \cdot \ldots \cdot a_{i_n}^{(n)}$.  Three-way tensors are formed by $\mathcal{X} = a \circ b \circ c$, as shown in Fig \ref{fig:3}a.

An essential operator of the Tucker decomposition is the $N$-mode product of a tensor with a matrix. Although $N$-mode products can be generalized to higher-order tensors, we focus only on the $N$-mode product between a tensor and a matrix since the Tucker decomposition uses it. The reader can check higher-order $N$-mode products and other types of tensor decomposition in \cite{Kolda:09}. Let $\mathcal{X}$ be a tensor in $\mathbb{R}^{I_1 \times I_2 \times \ldots \times I_n}$ and $\mathcal{U}$ a matrix in $\mathbb{R}^{J \times I_N}$. The $N$-mode product between $\mathcal{X}$ and $\mathcal{U}$ along its $N$ mode is denoted by $\mathcal{X} \times_{N} \mathcal{U}$ and has size $I_1 \times \ldots \times I_{N-1} \times J \times I_{N+1} \ldots I_{n}$. An entry in $\mathcal{X} \times_{N} \mathcal{U}$ is defined by:

\[ (\mathcal{X} \times_{N} \mathcal{U})_{i_1 \ldots i_{N-1} J i_{N+1} \ldots i_n} = \sum_{i_N=1}^{I_N} x_{i_1 i_2 \ldots i_n} \cdot u_{ji_N} . \]

The Tucker decomposition factorizes a tensor $\mathcal{X}$ in a core tensor $\mathcal{G}$ transformed by the $N$-mode tensor product along each of tensor's modes (dimensions). In the three-way case, where $\mathcal{X}$ in $\mathbb{R}^{I \times J \times K}$:

\[ \mathcal{X} \approx \mathcal{G} \times_{1} \mathbf{A} \times_{1} \mathbf{B} \times_{2} \mathbf{C} = \sum_{p=1}^{P} \sum_{q=1}^{Q} \sum_{r=1}^{R} g_{pqr} a_p \circ b_q \circ c_r,  \]

\noindent where $\mathcal{G} \in \mathbb{R}^{P \times Q \times R}$ is called the core tensor of the decomposition, and $\mathbf{A} \in \mathbb{R}^{I \times P}$, $\mathbf{B} \in \mathbb{R}^{J \times Q}$, and $\mathbf{C} \in \mathbb{R}^{K \times R}$ can be considered the main components of each mode of $\mathcal{X}$. $\mathcal{G}$ is dimensionally smaller than $\mathcal{X}$ ($P < I$, $Q < J$, and $R < K$) and can be viewed as a compressed version of $\mathcal{X}$, as we show in Fig \ref{fig:3}b. An element-wise Tucker decomposition entry is given by:

\[ x_{ijk} \approx \sum_{p=1}^{P} \sum_{q=1}^{Q} \sum_{r=1}^{R} q_{pqr} \cdot a_{ip} \cdot b_{jq} \cdot c_{kr}, \quad i \in 1 \ldots I, j \in 1 \ldots J, k \in 1 \ldots K. \] 

Balazevic \textit{et al.}~\cite{Balazevic:19} proposes a model for link prediction based on the Tucker tensor decomposition of the binary tensor representation of the knowledge graph. TuckER assumes that head and tail embeddings belong to the same vector space. Accordingly, TuckER defines an entity embedding matrix $\mathbf{E} \in \mathbb{R}^{n_e \times d_e}$, with $\mathbf{A} = \mathbf{C} = \mathbf{E}$. Relations are modeled using a relation embedding matrix $\mathbf{R} \in \mathbb{R}^{n_r \times d_r}$, which corresponds to the factor matrix $\mathbf{B}$ of the Tucker decomposition. Then, the tensor representation $\mathcal{X}$ of the knowledge graph is decomposed as:

\[ \mathcal{X} \approx \mathcal{G} \times_{1} \mathbf{E} \times_{2} \mathbf{R} \times_{3} \mathbf{E}. \]

The scoring function for triplet facts is defined for each triplet $\langle e_h, w_r, e_t \rangle$ using the element-wise tensor product per node:

\[ \phi(\mathbf{e_h}, \mathbf{w_r}, \mathbf{e_t}) = \mathcal{G} \times_{1} \mathbf{e_h} \times_{2} \mathbf{w_r} \times_3 \mathbf{e_t}, \]

\noindent where $\mathbf{e_h}$ and $\mathbf{e_t}$ are rows in $\mathbf{E}$ and $\mathbf{w_r}$ is a row in $\mathbf{R}$. TuckER applies a logistic sigmoid to the scoring function obtaining the predicted probability $p$ of a triplet being true.

For training, TuckER adds a reciprocal relation $\langle e_t, w_r^{-1}, e_h \rangle$ for every triplet fact $\langle e_h, w_r, e_t \rangle$. The training step uses a $1:N$ scoring, i.e., for a given triplet $\langle e_h, w_r, e_t \rangle$, TuckER scores $\langle e_h, w_r \rangle, \forall e_t \in \mathcal{E}$ and $\langle e_t, w_r^{-1} \rangle, \forall e_h \in \mathcal{E}$. Model fitting minimizes the Bernoulli log-likelihood loss function, that for a given triplet is defined by:

\[ \mathcal{L} = - \frac{1}{n_e} \sum_{i=1}^{n_e} \left( y^{(i)} \log (p^{(i)}) + (1 - y^{(i)}) \log (1 - p^{(i)}) \right), \]

\noindent where $y$ is the binary label vector of observed/unobserved facts in the knowledge graph.

\section{Results}

\subsection{Hyper-parameters tuning}

The best fit reported in S1 Table was trained using Adam~\cite{Kingma:15}, a method for first-order gradient-based optimization of stochastic objective functions, based on adaptive estimates of lower-order moments. This method is suitable for problems with sparse gradients. We used batch learning, with batch size = 128, learning rate = 0.005, and decay rate = 1.0. We choose all hyper-parameters by search based on validation set performance. The dimensionality of the relational space was tested with values in \{10, 20, 30\}. The best result was obtained for $d_r = 10$, which matches the number of relationships in SA19K. The dimensionality of the entity space was tested with values in \{100, 200, 500, 1000\}. The best result was obtained for $d_e = 200$. TuckER considers three dropout factors. The first operates on the input, which corresponds to the tensor representation of the entities. A second dropout factor applies to the tensor product between the tensor representations of the relationships and the Tucker core tensor. Finally, a third dropout factor is applied when calculating the tensor product between the previous tensor and the tensor representation of the entities. This final product corresponds to the tensor representations of the pair $\langle s, r \rangle$, which is used for the 1:N ranking-based training strategy. These parameters were evaluated using grid search in \{0.2, 0.3, 0.4, 0.5\}. The best configuration corresponds to factors \{0.5, 0.2, 0.2\}. Balazevic \textit{et al.}~\cite{Balazevic:19} comment that low dropout factors are required for datasets with a high number of training triplets per relation, avoiding the risk of over-fitting. Our high input dropout value suggests that the number of training triplets is low regarding the number of entities available in SA19K. However, the proportion of examples per relationship is high due to the low number of relations in SA19K, which explains the lowest values obtained for the second and third dropout factors. 

\subsection{Performance per income decile}

\begin{figure}
    \centering
    \begin{subcaptionblock}{.19\columnwidth}
        \includegraphics[width=\columnwidth]{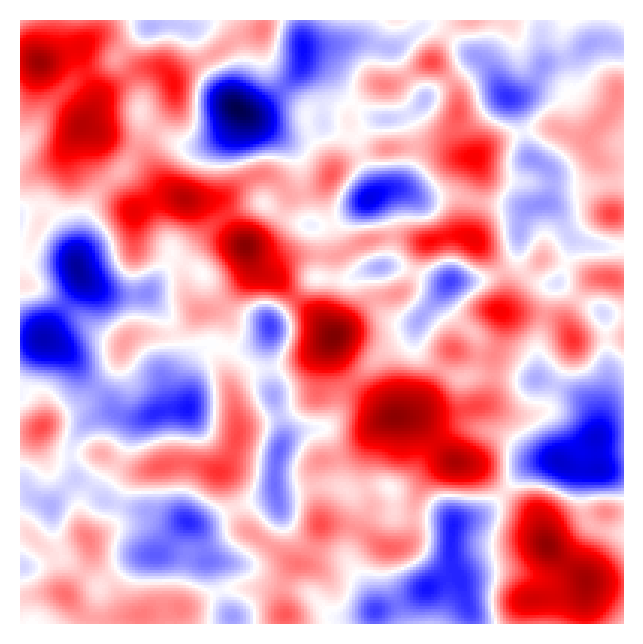}
        \caption{D1.}
    \end{subcaptionblock}
    \begin{subcaptionblock}{.19\columnwidth}
        \includegraphics[width=\columnwidth]{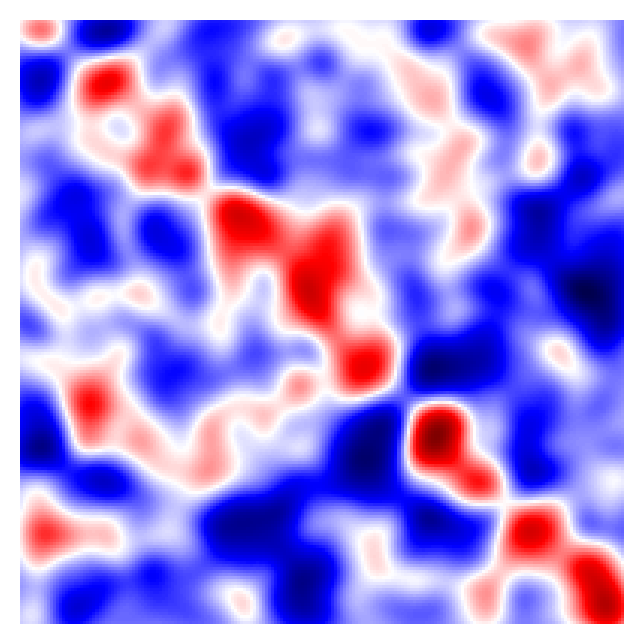}
        \caption{D2.}
    \end{subcaptionblock}
    \begin{subcaptionblock}{.19\columnwidth}
        \includegraphics[width=\columnwidth]{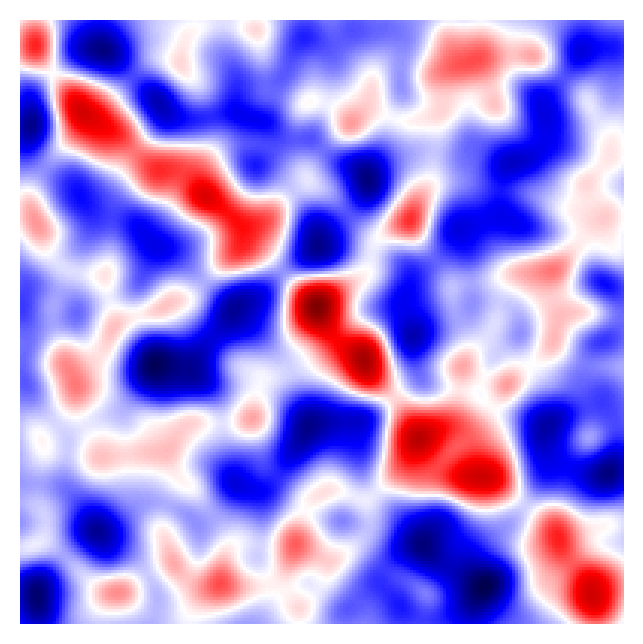}
        \caption{D3.}
    \end{subcaptionblock}
    \begin{subcaptionblock}{.19\columnwidth}
        \includegraphics[width=\columnwidth]{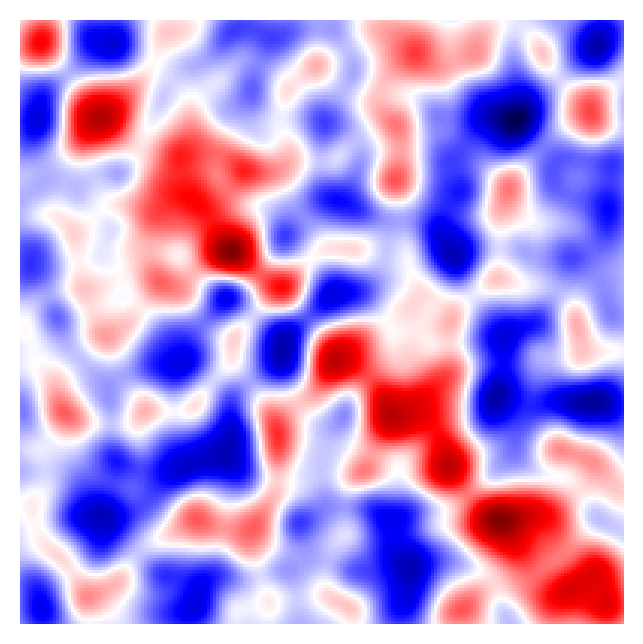}
        \caption{D4.}
    \end{subcaptionblock}
    \begin{subcaptionblock}{.19\columnwidth}
        \includegraphics[width=\columnwidth]{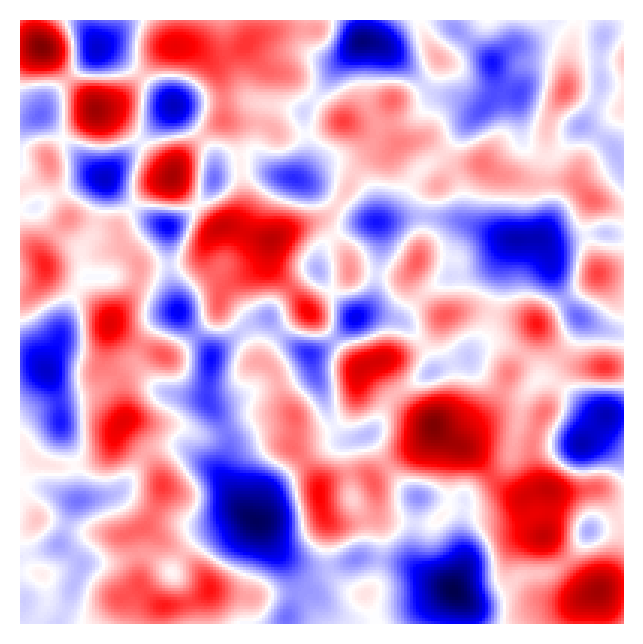}
        \caption{D5.}
    \end{subcaptionblock}
    \begin{subcaptionblock}{.19\columnwidth}
        \includegraphics[width=\columnwidth]{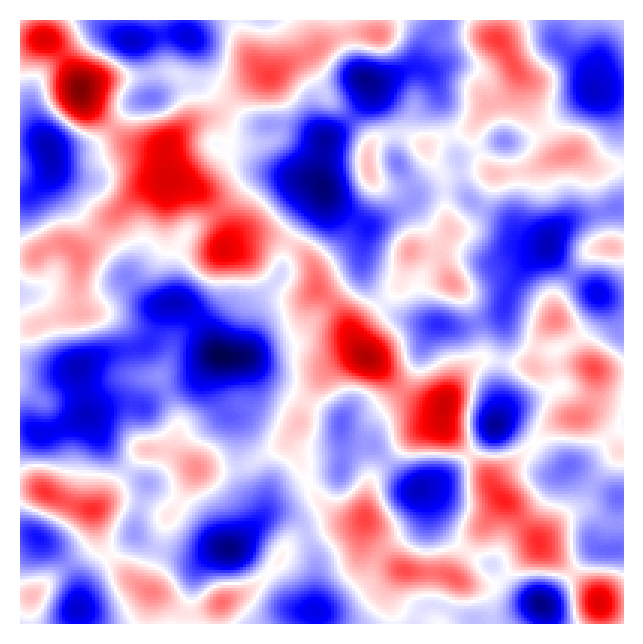}
        \caption{D6.}
    \end{subcaptionblock}
    \begin{subcaptionblock}{.19\columnwidth}
        \includegraphics[width=\columnwidth]{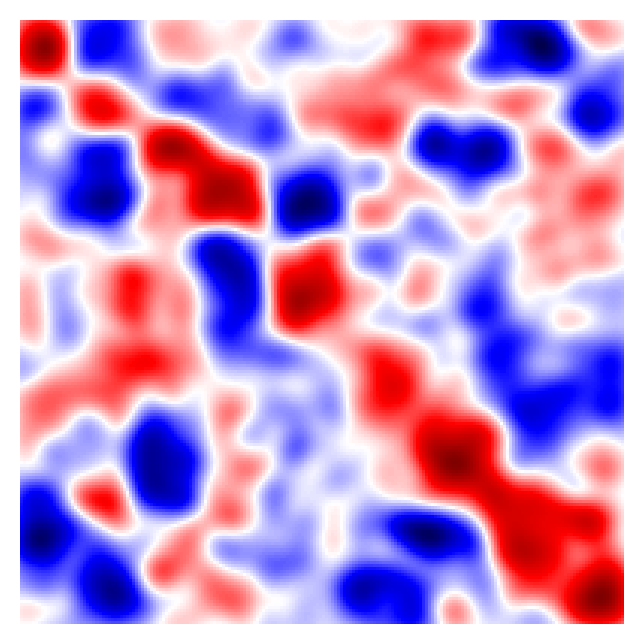}
        \caption{D7.}
    \end{subcaptionblock}
    \begin{subcaptionblock}{.19\columnwidth}
        \includegraphics[width=\columnwidth]{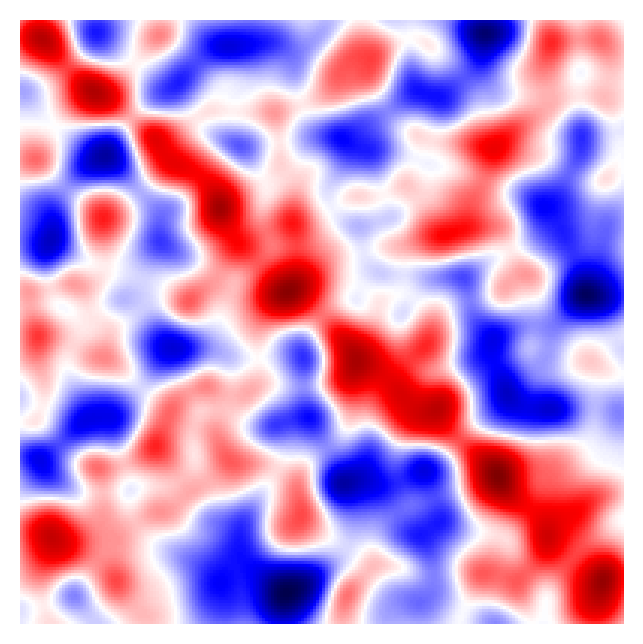}
        \caption{D8.}
    \end{subcaptionblock}
    \begin{subcaptionblock}{.19\columnwidth}
        \includegraphics[width=\columnwidth]{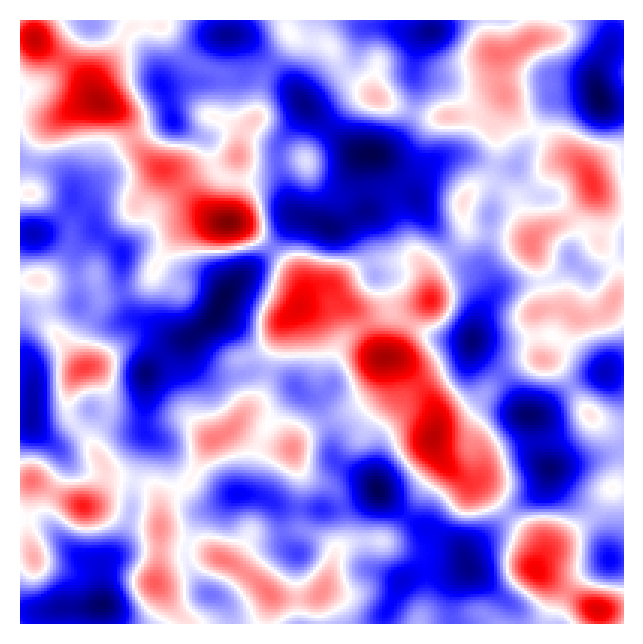}
        \caption{D9.}
    \end{subcaptionblock}
    \begin{subcaptionblock}{.19\columnwidth}
        \includegraphics[width=\columnwidth]{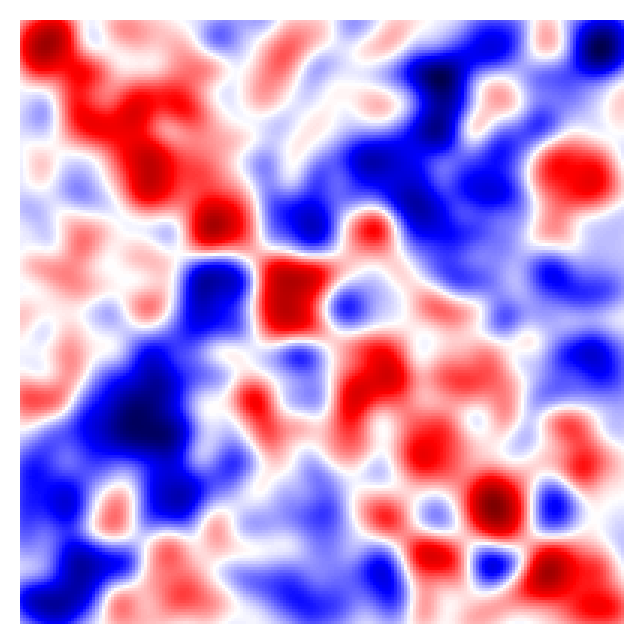}
        \caption{D10.}
    \end{subcaptionblock}
    \caption{\textbf{Heatmaps of the relation matrices learned with TuckER for each income decile.} Red and blue colors indicate high and low-value entries in the relation matrices, respectively. These matrices are computed using $k$-mode tensor products between $\matr{R}$ and the second mode of the core tensor $\matr{W}$. Although the model does not force the learning of symmetric relationships, all the learned relationships are approximately symmetric, indicating a good fit of the model to SA19k.}
    \label{fig:4}
\end{figure}

Fig \ref{fig:4} shows the relation matrices learned by TuckER for each income decile. Although TuckER does not force the learning of symmetric relationships, the learned matrices are approximately symmetric, suggesting a good fit of the model to SA19k. Heatmaps with more red entries indicate that TuckER encodes more information in that relationship than in the rest of the deciles. Fig \ref{fig:4} shows that TuckER encodes many information in relationships D1 and D5, which are the ones with the most burned entries in the tensor. On the other hand, D2-D4 show fewer burned entries in the tensor, suggesting that these relationships are less informative than the others. SA19K shows that low and middle-income deciles have less diverse relationships between surnames, with a low incidence of affinity ties in this network segment. The effort made by TuckER in D1 is related to the low incidence of ties in this layer (see Fig \ref{fig:1} (b)). In D5, this income decile articulates the low-income deciles (D1-D5) with the middle and high-income deciles (D6-D10), suggesting a transition in SA19K. The relation matrices D6-D9 obtained by TuckER are quite similar to each other, showing homogeneity in this income range in terms of surname affinity. In D10, TuckER shows more burned entries in the tensor diagonal, which coincides with the high incidence of affinity ties in this segment of the network (see Fig \ref{fig:1} (b)).

For each triplet in the test partition, TuckER evaluates the pair $\langle s, r \rangle$ with the rest of the network entities, using a 1:N ranking-based strategy. Accordingly, for the 19041 triplets evaluated, TuckER orders the triplets in descending order according to the softmax layer's logit. Then, the ordered list of triplets is evaluated by counting hits@n. A grounded triplet is a hit@n if it is among the first n results with the model's highest logit. It should be noted that the problem is very hard since the objective is to rank grounded triplets in the top positions of a long list of possible ties. According to the degree distribution in SA19K, on average, each entity has ~20 affinity ties. Since the length of each entity list is 19041, the chances of generating a top@n list of maximum precision at random are $\Pi_{i=1}^{20} \frac{21-i}{19041-i} \approx 0.00105^{20}$, a extremely small number. TuckER, on average, gets hits@1 = 0.214, hits@3 = 0.406, and hits@10 = 0.607, as shown in S1 Table. These results indicate that surname affinity ties are strongly predictable. 

\begin{figure}
    \centering
    \begin{subcaptionblock}{.49\columnwidth}
        \includegraphics[width=\columnwidth]{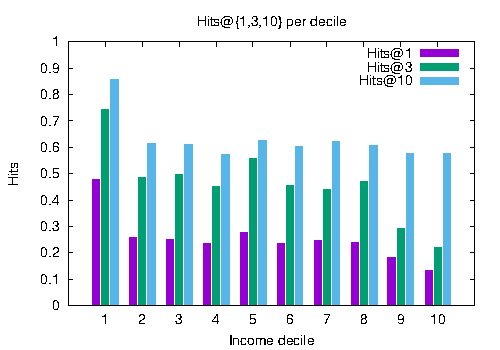}
        \caption{Hits@1, 3, and 10 per income decile.}
    \end{subcaptionblock}
    \begin{subcaptionblock}{.49\columnwidth}
        \includegraphics[width=\columnwidth]{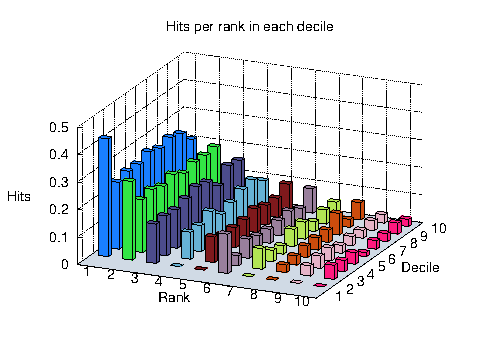}
        \caption{Hits per rank in each income decile.}
    \end{subcaptionblock}
    \caption{\textbf{TuckER manages to rank high many grounding triplets, which allows the model to perform well on hits @1, @3, and @10;} a) While the results in hits@1 show a better performance in the lower deciles, in hits@10, the results are more even; b) When disaggregating the results in absolute hits per rank position, we see that TuckER does not have significant differences between deciles, placing the vast majority of hits in the first five positions of each list.}
    \label{fig:5}
\end{figure}

Fig \ref{fig:5}a shows hits @1, @3, and @10 for each SA19k income decile in the test partition. While decile 1 is the one with the highest predictability, decile 10 is the least predictable. Between deciles 2 and 10, all hits @10 reach around 0.6. These results are related to the diversity of surnames in each socioeconomic segment. In \cite{Bro:21}, we showed that the diversity of surnames, quantified according to the alpha coefficient \cite{Barrai:01}, is much greater in high socioeconomic segments than in low segments. Affinity tie prediction results indicate that the higher the surname diversity, the more difficult it is to predict the affinity ties. Regarding the precision of the method, the greatest variations are found in hits@1, where the affinity of surnames becomes less predictable as the income decile increases. Fig \ref{fig:5}b shows the results disaggregated at the rank-lavel, i.e., the absolute number of hits by ranking position. We can see that the hits decrease as the position rank increases, which indicates that the TuckER logits help to rank in the top positions the triplets with the highest probability of generating affinity ties. Most of the hits are in the top five ranking positions.

\subsection{Explaining surname affinity predictions}

We evaluate the ability of our model to forecast the emergence of new triangles. We consider three sources of information. The first source consists of the original network, which we call grounded triplets. If two unconnected nodes have many neighbors in common, a quantifiable measure known as the shared nearest neighbors fraction (SNN), the model expects them to connect. SNN measures the fraction of neighbors in common over the union of the neighborhoods of two entities. SNN is a very relevant measure in clustering analysis~\cite{Zamora:16}, since it reveals clues for high-dimensional data clustering of varying density. In social networks, SNN relates to the friend of a friend mechanism (FOAF), which explains that two strangers have a high probability of engaging if they have friends in common. To account for this effect, we measure the SNN between the source and target entity of each new grounded triplet in the same income decile. The second source of information is the SNN fraction in nearest income deciles. If two unknown nodes share many neighbors in neighboring income deciles, the model expects them to connect. We estimate this effect by counting the fraction of SNN in the same, and the two nearest deciles. The third source of information is the neighborhood in the embedding space. We measure this effect by counting the fraction of SNN in the embedding space, according to the k-mode tensor product of the entities of the triplet along with the relation matrix that represents each income decile. To measure the SNN in the embedding space, we compute a kNN query using the Euclidean distance with $k=50$, for each entity embedding of a grounded predicted triplet. If the neighbors of two disconnected nodes intersect in the embedding space (i.e., both entities show a high fraction of SNN), the model expects them to connect.

\begin{figure}
    \centering
    \begin{subcaptionblock}{.49\columnwidth}
        \includegraphics[width=\columnwidth]{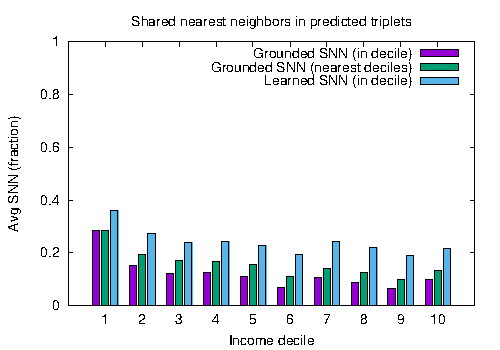}
        \caption{SNN in predicted grounded triplets.}
    \end{subcaptionblock}
    \begin{subcaptionblock}{.49\columnwidth}
        \includegraphics[width=\columnwidth]{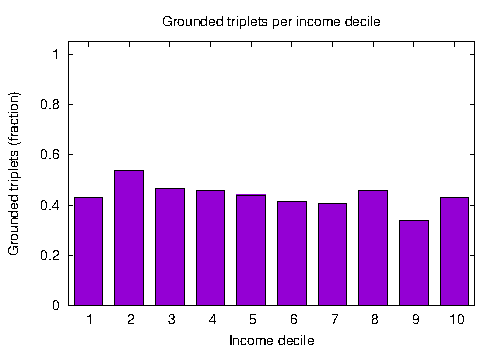}
        \caption{Grounded triplets per income decile.}
    \end{subcaptionblock}
    \caption{\textbf{Shared nearest neighbors (SNN) and grounded triplets per income decile.} a) The fraction of SNN (average) varies depending on which source of information we consider. The most informative source is proximity-based SNN, although there is significant support based on grounded SNN from the original network. b) Grounded triplets per income decile. The fraction of grounded triplets is always less than half of the predicted triplets, except in decile 2, where it reaches almost 60\% of the predictions}
    \label{fig:6}
\end{figure}

\begin{figure}
    \centering
    \begin{subcaptionblock}{.49\columnwidth}
        \includegraphics[width=\columnwidth]{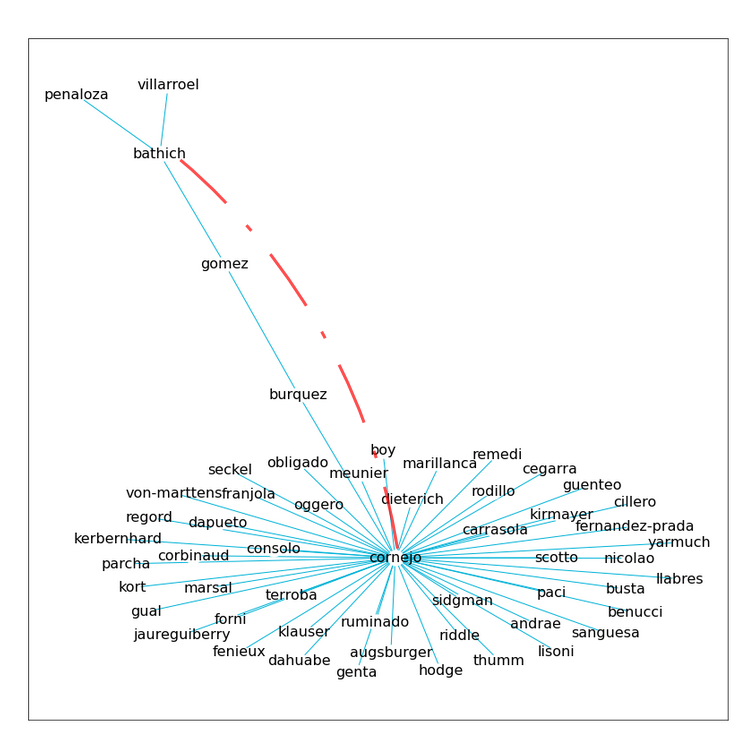}
        \caption{('bathich', 'd10', 'cornejo').}
    \end{subcaptionblock}
    \begin{subcaptionblock}{.49\columnwidth}
        \includegraphics[width=\columnwidth]{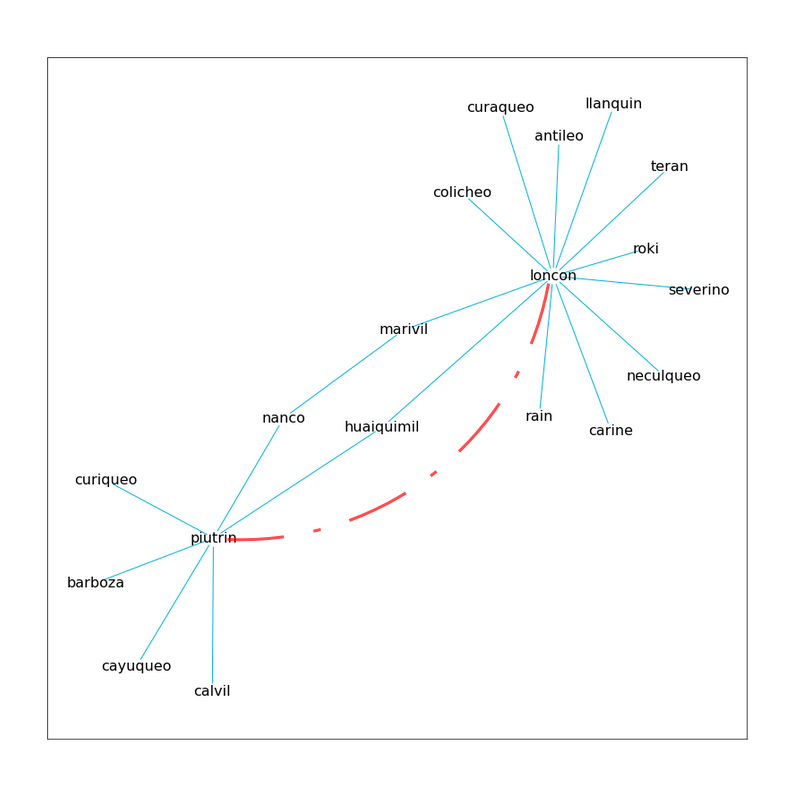}
        \caption{('loncon', 'd9', 'piutrin').}
    \end{subcaptionblock}
    \begin{subcaptionblock}{.49\columnwidth}
        \includegraphics[width=\columnwidth]{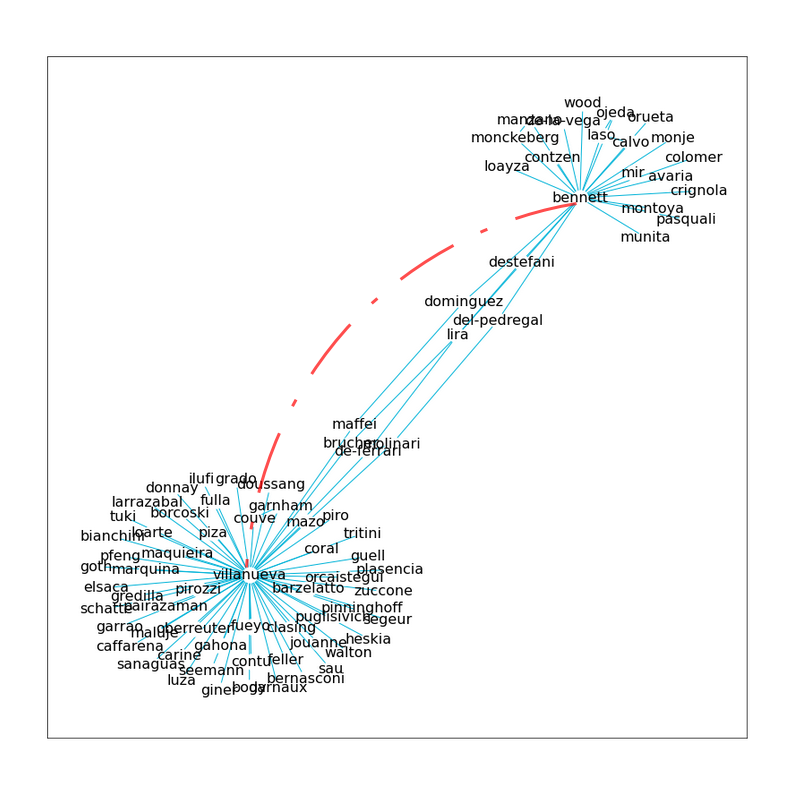}
        \caption{('bennett', 'd10', 'villanueva').}
    \end{subcaptionblock}
    \begin{subcaptionblock}{.49\columnwidth}
        \includegraphics[width=\columnwidth]{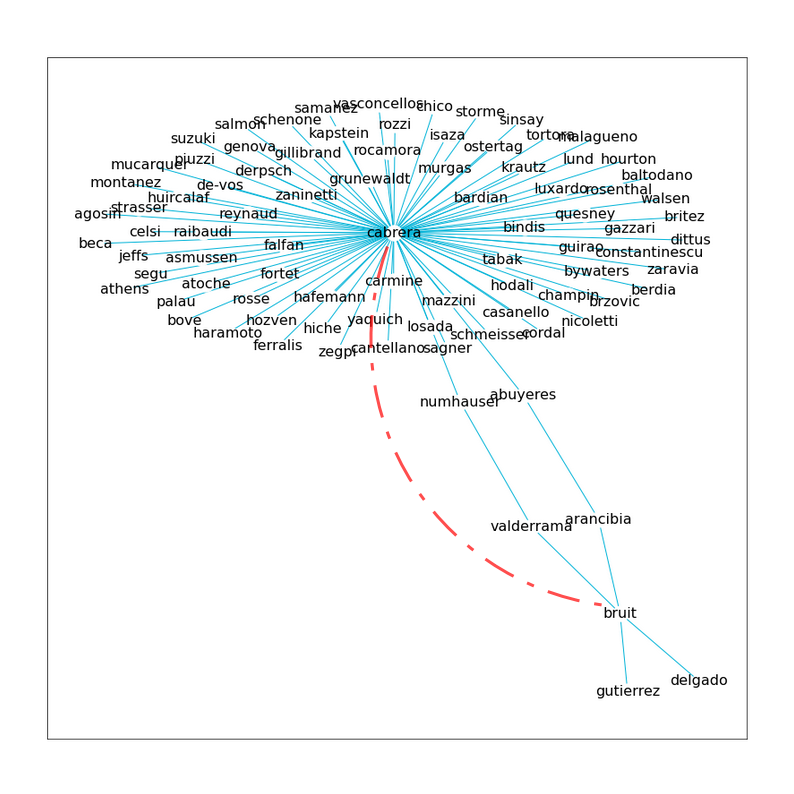}
        \caption{('bruit', 'd10', 'cabrera').}
    \end{subcaptionblock}
    \caption{\textbf{Grounded predictions generally show paths that connect two micro-clusters.} Figures a and d show the connection of a larger cluster and a smaller one, exhibiting an absorption pattern. Figures b and c show the relationship of two clusters of similar sizes. In all the cases indicated, the new connection (marked with a red dotted line) creates a shorter path between both micro-clusters, decreasing the resulting graph's diameter.}
    \label{fig:7}
\end{figure}

Fig \ref{fig:6}a shows that the SNN fraction varies by source of information. The lowest fraction of SNN in predicted triplets is obtained from grounded triplets of the same decile. This pattern is more robust in the lowest income deciles, losing relevance as the socioeconomic level increases. After adding near income deciles (we use the two closest deciles in the evaluation), the fraction of SNN increases in all deciles. This pattern shows that this second source of information is more informative than the first. Finally, after adding friends using neighbors in the embedding space, the fraction of SNNs increases. This finding shows that the node embedding space allows capturing ungrounded shared neighbors, representing valuable information for all income deciles. The embedding space is the most informative dimension for the affinity prediction task. In specific, this space is very relevant in income deciles 1, 2, and 7. In fact, in all income deciles, the grounded predicted triplets share more neighbors in this space than in the empirical network. This finding suggests that the entity embedding space can encode high-order information that is different to the one encoded from the neighborhoods of the empirical network.

Fig \ref{fig:7} shows examples of grounded predictions. All of them are characterized by connecting two micro-clusters adding an interaction between them. In some cases (Figures \ref{fig:7}a and \ref{fig:7}d), we observe an absorption pattern, where the interaction occurs between micro-clusters of very different sizes. In other cases, the interaction occurs between clusters of similar sizes (Figures \ref{fig:7}b and \ref{fig:7}c). In all these cases, the new interaction (marked with a red dotted line) creates a new path between both clusters, decreasing the resulting graph's diameter.

Most of the predicted triplets (57\% on average across deciles) come from micro-clusters without direct connections like those shown in Fig \ref{fig:7}. TuckER manages to learn proximity relationships between nodes obtained by working with the tensor factorization of the entire network. This principle allows identifying long-term interaction patterns between unconnected nodes, which cannot be detected by locality-based methods such as those based on FOAF patterns. Fig \ref{fig:6}b shows the fraction of grounded predictions per income decile that corresponds to high SNN fractions in the same decile and/or in close income deciles, which on average represents the 43\% of the successfully predicted triplets. Fig \ref{fig:6}b shows that more than half of the predicted triplets are grounded on the entity embedding space. Accordingly, TuckER must make a significant effort to identify triplets based on proximity-based neighbors inferred from the entity embedding space, representing more than half of the correctly predicted triplets. 

\section{Discussion}

The results show great predictability in the formation of ties between people holding certain surnames. Triadic closure is an empirical regularity well established in the social sciences ~\cite{deutschlandSociologyInquiriesConstruction2009}. If persons A and B share a common friend C, they are much likelier to become friends than in the alternative scenario where they do not share C as a friend. So far, most research has examined triadic closures either theoretically ~\cite{granovetter73, burtStructuralHolesSocial1995}, to statistically model networks ~\cite{snijdersIntroductionStochasticActorbased2010}, or in micro-empirical contexts. This paper empirically investigates the formation of triads at the level of an entire population. It shows that surname-based groups are prone to form triads.

Chile is one of the most unequal countries in the world~\cite{OECD:01}, with a pattern of social mobility where it is especially hard to enter the elites. \cite{torcheUnequalFluidSocial2005}. Our previous research~\cite{Bro:21} showed that Santiago's upper class is highly endogamous, not unlike an ethnic minority in its tendency to form in-group marital unions. This pattern can be explained in two ways: from the preferences of wealthy individuals to connect with others similar to them, or from the structure of the network around them.

The results of this study reveal the importance of the second mechanism, a dimension of socioeconomic segregation that is independent from homogamy: the subjective preference that people may have to connect with others like themselves~\cite{mcphersonBirdsFeatherHomophily2001}. People also connect with and marry their socioeconomic in-groups because the latter inhabit the same locations in the network. In this sense, an upper-class person may not have a particular preference of marrying someone of her own status group, but she may end up doing so regardless because the network of social interactions constrains the marital options available to her.

Importantly, we make a distinction between empirical and more abstracts forms of network proximity. Neighbors may come in the form of grounded triplets, but also as proximate points in the embedding space. We show that triadic closure occurs not only in the empirical network, but also in the embedding space. If nodes A and B are both similar to C, they have a good chance of forming a tie in the network. Link prediction tasks therefore would benefit from taking not only the empirical graph into account but also the embeddings of nodes.

\section{Conclusion}

This paper develops a model to predict the formation of new links in the network of surname affinity in Santiago, Chile (SA19k). It formulates the problem as a knowledge base completion task, and finds that the TuckER method produces the best results. The results show that the formation of affinity links between surnames in Santiago, Chile, is highly predictable. We find that proximity in the network explains a large proportion of new links. Importantly, the method distinguishes between proximity in the empirical network and proximity in the embedding space, and finds that the latter is more predictive of link formation than the former.

The results shed light on an unexplored dimension of socioeconomic segregation. People may or may not prefer connecting with and marrying others belonging to their same socioeconomic status. However, given that the network of social interactions places constraints on their options, they are still likely to connect with or marry in-groups. Extensions of this paper could incorporate proximity in the embedding space as a parameter in statistical models that aim to disentangle the effects of different types of network proximity in explaining the social segregation that we often observe in empirical networks.

\section*{Acknowledgements}

Thanks to Josué Tapia and Andrés Cruz for helping in the data-building phase of this project. Other persons that helped are Daniel Alcatruz, Sebastián Huneeus, and Johans Peña. 

\bibliographystyle{unsrtnat}
\bibliography{template}

\newpage

\appendix

\section*{Appendix}

\begin{table}[h]
    \centering
    \caption*{\textbf{S1 Table. Link prediction for SA19k. The results show the performance of the models in hits@\{1,3,10\}, and Mean Reciprocal Rank (MRR).}}
    \vspace{4mm}
    \begin{tabular}{|l|c|c|c|c|} \hline
    
    Model                       &  H@1   & H@3   & H@10  & MRR   \\ \hline \hline
    TransE      &  0.073 & 0.358 & 0.654 & 0.263 \\
    TransH        &  0.068 & 0.275 & 0.497 & 0.213 \\
    DistMult      &  0.061 & 0.115 & 0.223 & 0.115 \\
    ComplEx  &  0.085 & 0.136 & 0.226 & 0.134 \\
    ConvE     &  0.193 & 0.381 & 0.593 & 0.325 \\
    RotatE         &  0.181 & 0.388 & \textbf{0.668} & 0.332 \\
    CrossE       &  0.153 & 0.351 & 0.610 & 0.229 \\
    ConvTransE   &  0.171 & 0.324 & 0.488 & 0.278 \\
    SACN         &  0.156 & 0.275 & 0.408 & 0.243 \\
    TuckER   &  \textbf{0.214} & \textbf{0.406} & 0.607 & \textbf{0.346} \\
    \hline
    \end{tabular}
\end{table}

\end{document}